\documentclass{aa}

\usepackage{graphicx}
\usepackage{xcolor}
\usepackage[position=top]{subfig}
\usepackage[fleqn]{amsmath}
\usepackage[utf8]{inputenc}
\usepackage{epsfig, amsmath, amssymb}
\usepackage[varg]{txfonts}
\usepackage{multirow}
\usepackage{tcolorbox}

\definecolor{myblue}{rgb}{0.00, 0.0, 0.9}
\definecolor{myred}{rgb}{0.90, 0.0, 0.0}
\definecolor{mygreen}{rgb}{0.0, 0.7, 0.0}
\usepackage[breaklinks, citecolor=myblue, linkcolor=myred, urlcolor=purple, colorlinks=true, debug, baseurl=' ']{hyperref}

\titlerunning{High-Frequency QPO in NGC 1365}
\authorrunning{Y. K. Yan et al.}

\begin{document}

\title{An X-ray high-frequency quasi-periodic oscillation in NGC 1365}

\author{
Y. K. Yan\inst{1}, P. Zhang\inst{1, 2}\thanks{E-mail:zhangpeng@ctgu.edu.cn}, Q. Z. Liu\inst{3}, Z. Chang\inst{4}, G. C. Liu\inst{1, 2}\thanks{E-mail:gcliu@ctgu.edu.cn}, J. Z. Yan\inst{3}, X. Y. Zeng\inst{1, 2}}
\institute{
College of Science, China Three Gorges University, Yichang 443002, China
\and
Center for Astronomy and Space Sciences, China Three Gorges University, Yichang 443002, China
\and
Purple Mountain Observatory, Chinese Academy of Sciences, Nanjing 210008, China
\and
Key Laboratory of Particle Astrophysics, Institute of High Energy Physics, Chinese Academy of Sciences, 19B Yuquan Road, Beijing 100049, China
}

\date{Received XXX / Accepted XXX}

\abstract{
This study presents the detection of a high-frequency quasi-periodic oscillation (QPO) in the Seyfert galaxy NGC 1365 based on observational data obtained by \emph{XMM-Newton}  in January 2004. Utilizing the weighted wavelet Z-transform (WWZ) and Lomb-Scargle periodogram (LSP) methods, a QPO signal is identified at a frequency of $2.19 \times 10^{-4}\ {\rm Hz}$ (4566 s), with a confidence level of 3.6\ $\sigma$. The signal is notably absent in the lower 0.2 -- 1.0 keV energy band, with the primary contribution emerging from the 2.0 -- 10.0 keV band, where the confidence level reaches 3.9 $\sigma$. Spectral analysis shows that there are multiple absorption and emission lines in the high-energy band (> 6\ keV). The correlation between the QPO frequency ($f_{\rm QPO}$) and the mass of the central black hole ($M_{\rm BH}$) of NGC 1365  aligns with the established logarithmic trend observed across black holes, indicating the QPO is of high frequency. This discovery provides new clues for studying the generation mechanism of QPOs in Seyfert galaxies, which helps us understand the accretion process around supermassive black holes and the characteristics of strong gravitational fields in active galactic nuclei.
}


\maketitle

\section{Introduction}
\label{sec:intro}

Active galactic nuclei (AGNs) are a class of extragalactic objects characterized by intense energetic processes occurring at their cores, which is primarily driven by the accretion of matter onto a supermassive black hole (BH). The presence of anisotropic structures within the galactic nuclei bestows a variety of observational characteristics across the electromagnetic spectrum \citep{Swain2023}, with the X-ray spectrum notably shaped by the reradiation resulting from interactions between X-ray photons and the surrounding gas. This X-ray radiation is believed to originate from regions within the accretion disk or closer to the black hole \citep{Ghisellini1994}. A defining feature of AGN X-ray emission is its temporal variability, which can be attributed to changes in the accretion processes and the shielding effects of the environment \citep{Lachowicz2005}. Quasi-periodic oscillations (QPOs) are a special phenomenon of some X-ray and $\gamma$-ray emission sources, and are quite common in neutron stars and BH binaries in the Milky Way and nearby galaxies \citep{Gierlinski2008, Pan2016}. However, this phenomenon is very rare in AGNs \citep{Zhang2020}, and many early detections have been disfavored due to inadequacies in modeling the underlying broadband noise \citep{Alston2015}.

An important observational property of stellar-mass BH X-ray binaries, potentially a probe of the inflow structure just outside the horizon, is the presence of QPOs \citep{Remillard2006}. These phenomena exhibit different peaks in their X-ray power density spectrum. QPOs are usually divided into high frequency (HF; $\sim$10 -- $10^{3}$ Hz) and low frequency (LF; $\sim$ 10$^{-2}$ -- 10 Hz). The latter are further classified into subtypes based on their coherence and the intensity of different frequency bands (Type A, B, and C: \citealt{Casella2004}). Coherence is represented by the quality factor Q $= \nu/\Delta\nu$, where $\nu$ is the centroid frequency and $\Delta\nu$ is the full width at half maximum near the centroid frequency. The most common QPO subtype (and also the subtype with the highest Q-factor) is type C; its frequency usually ranges from a few mHz to 10 Hz, but sometimes frequencies as high as 30 Hz can be detected \citep{Revnivtsev2000}. The explanation for the origin of type C QPO is still controversial; alternative models include internal disk instability and Lense Thirring (LT) precession of jets or internal heat flux or corona \citep{Stella1999, Ingram2009, Ingram2011}.

These QPO phenomena exhibit similarities across compact objects of different masses and types, such as microquasars, AGNs, or intermediate-mass BHs. The correlation between HF QPOs and LF QPOs observed in neutron stars may also exist in weak galactic BH QPO data (see \citealt{motta2022}). Although multiple models have attempted to explain the origin of these QPOs, no unified theory has yet been established \citep{Zhang2020}. Explanations for the origin of different types of QPOs vary significantly. For LF QPOs, \cite{Stella1999} proposed the so-called relativistic precession model, which links QPOs to the radial (periastron) and vertical (LT) precessions of inhomogeneities in the accretion disk, approximated as test particles. The origin of HF QPOs, on the other hand, may be related to the orbital motion of the accretion flow within the disk, with relatively stable frequencies that do not significantly change with luminosity. The main explanatory models include the resonance model \citep{Abramowicz2001}, accretion-ejection instability within the disk \citep{Tagger1999}, and disk oscillations in different modes (g, c, and p mode) \citep{Wagoner1999}.

Furthermore, the relationship between BH mass (M) and QPO frequency  is of significant importance in the present study, as is its potential connection with the relativistic 1/M scaling of orbital frequencies. The expected 1/M scaling is generally valid for most orbital models, with recent key findings supporting this evidence \citep{Goluchova2019}. Additionally,  extensive studies have shown that QPOs typically exhibit transient characteristics, indicating a short-lived phenomenon \citep{Gierlinski2008, Pan2016}. Regardless of these rapidly changing mechanisms, they occur very close to the BH, which exerts the most direct influence on them. By studying the characteristics of QPOs, we can gain a better understanding of the accretion processes from stellar-mass to supermassive BHs and the accretion disk theory surrounding them \citep{Zhang2020}. This also provides a critical way to study the strong gravitational fields around BHs.

NGC 1365 is a large barred spiral galaxy in the Fornax galaxy cluster \citep{sandage1983}, with a distance of about 18.6 $\pm$ 0.6 Mpc \citep{Madore1999, Silbermann1999, Springob2009, Polshaw2015, Jang2018}. According to the latest calculations, its redshift value is z = 0.0051 \citep{Koss2022}. This spiral galaxy has been extensively investigated (a review of early work can be found in \citealt{Lindblad1999}). The galaxy contains an active core, and its spectral emission lines have both broad and narrow lines. The mass of the central BH is in the range of (5 -- 10) $ \times 10^{6}\ M_{\odot}$ \citep{Fazeli2019}, and the mass of the supermassive BH can be derived from the emission lines in the broad-line region, in particular the broad components of hydrogen recombination lines, which can be calculated using the formula obtained in the near-infrared (NIR) region \citep{Kim2010}. NGC 1365 is also a typical Seyfert galaxy, and according to the Seyfert classification, it belongs to type 1.8 \citep{Maiolino1995}. However, some scholars have classified it as type 1, 1.5, and 2 (e.g., \citealt{Veron1980, Turner1993, Risaliti2007, Thomas2017}). This is due to the different classifications caused by the change in X-ray flux and the absorption body density along the line of sight \citep{Hernandez2015}. The X-rays from the core include the hard continuous radiation of the active nucleus itself, the Fe-k line radiation of the rotating disk, and the thermal radiation of the surrounding stellar burst activity.

In this work, we employ the weighted wavelet Z-transform (WWZ) and the Lomb-Scargle periodogram (LSP) to detect and characterize a QPO signal in the X-ray observations of NGC 1365. Our findings provide insights into the ongoing discussion regarding the transient nature of QPOs and their potential correlation with the properties of the central BH. The structure of this paper is as follows: Section 2 details our observations, data reduction, and analysis methods, while in Section 3 we summarize our results and discuss their implications for the study of AGNs and for the broader astrophysics community.

\section{Observations and data analysis}
\subsection{Observations and data reduction}
\label{sec:Observations}
\begin{figure*}[!h]
        \centering
        \includegraphics[scale=1]{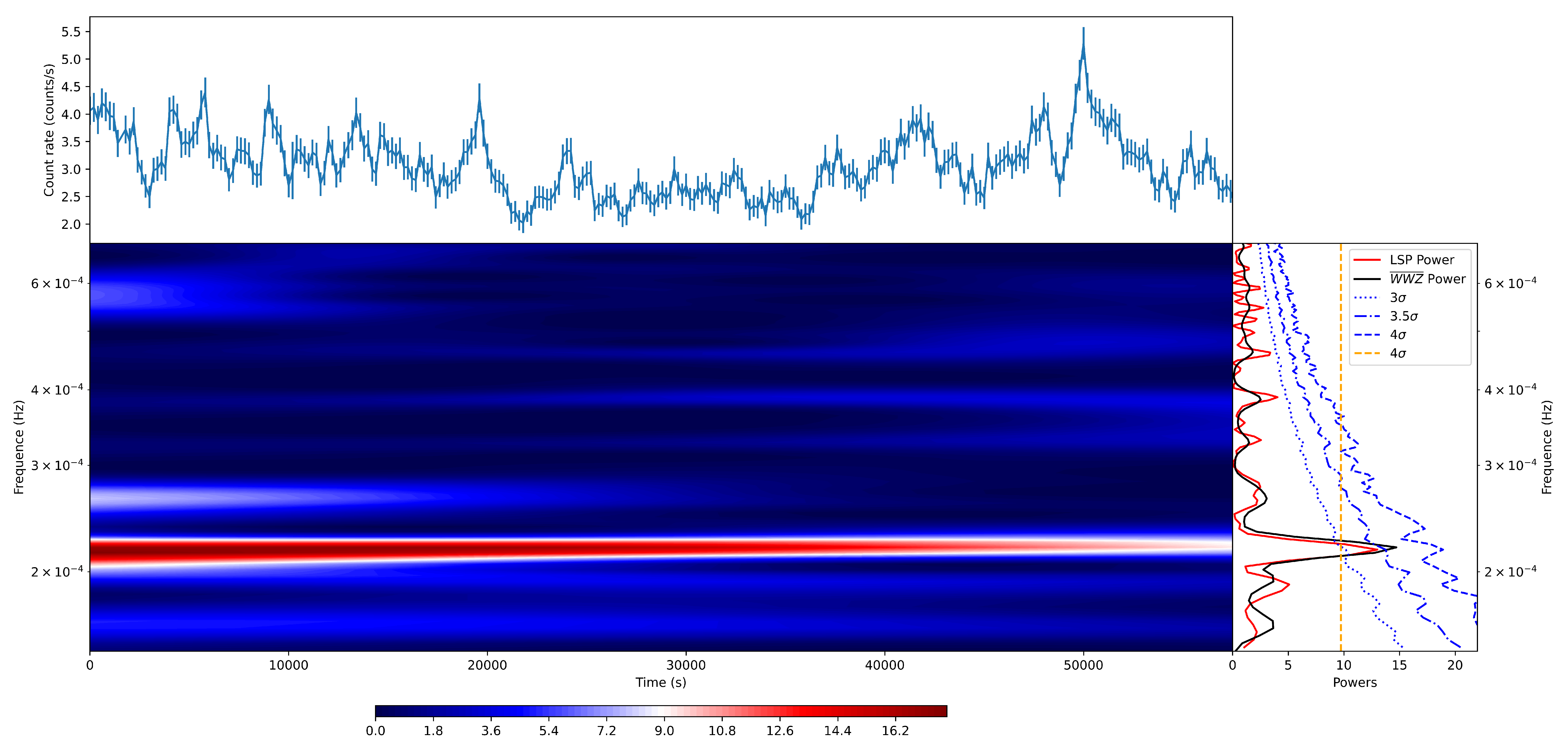}
        \caption{
		Light curve and WWZ power of EPIC camera of the \emph{XMM-Newton} ObsID 0205590301. The upper panel shows the light curve in the range of 0.2 -- 10.0\ keV.
        The upper panel shows the light curve of the EPIC camera of the  ObsID 0205590301 in the range of 0.2 -- 10.0\ keV. The lower-left panel displays a two-dimensional plot of dynamic WWZ power in time--frequency space. The lower-right panel displays average WWZ power (black) and LSP value (red) curves. The blue lines represent the confidence levels of 3.0, 3.5, and 4.0\ $\sigma$, respectively. The orange dotted line represents the 4.0\ $\sigma$ confidence level calculated using the method proposed by \citet{baluev2008}.
        }
        \label{one}
\end{figure*}
\begin{figure}[!h]
\centering
\includegraphics[scale=0.55]{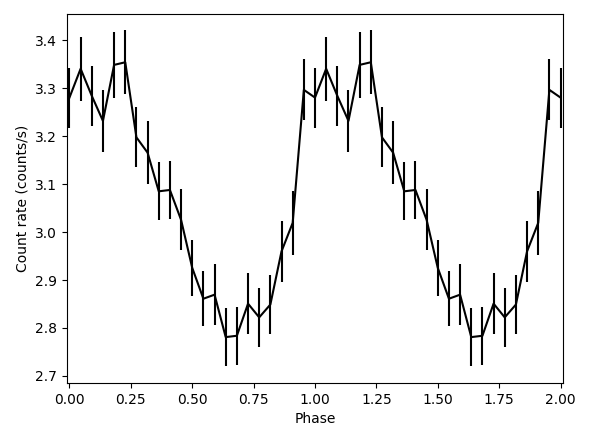}
\caption{
Pulse shape of the light curve, folded using a period of 4566 s, with two cycles shown.
}
\label{two}
\end{figure}
\begin{figure}[!h]
        \centering
        \includegraphics[scale=0.27]{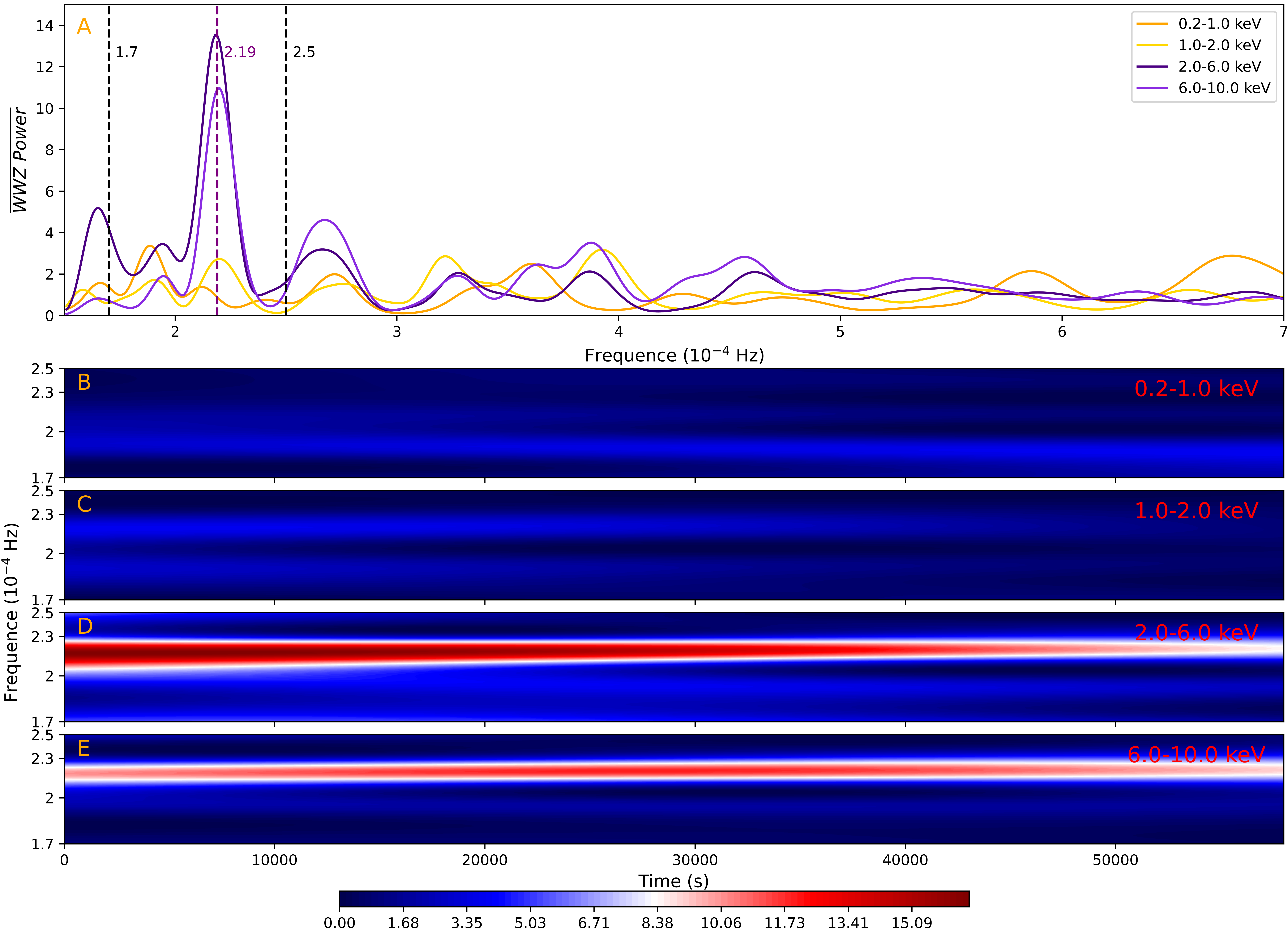}
        \caption{
		WWZ power for ObsID 0205590301 at different energy band. Panel A shows average WWZ power for ObsID 0205590301 at 0.2 -- 1.0\ keV, 1.0 -- 2.0\ keV, 2.0 -- 6.0\ keV, and 6.0 -- 10.0\ keV. Panels B, C, D, and E represent the results of wavelet analysis at different energy bands, with the frequency range between two black dashed lines in Panel A.
        }
        \label{three}
\end{figure}
\emph{XMM-Newton} is an X-ray detection satellite launched by the European Space Agency on December 10, 1999. It carries two sets of X-ray detectors, including three European Photon Imaging Cameras (EPIC: PN, MOS1, and MOS2; \citealt{Strder2001, Turner2001}) and two Reflection Grating Spectrometers (2RGS; \citealt{Herder2001}). Within the span of 2003 and 2004 alone, \emph{XMM-Newton} conducted six observations of NGC 1365. After a considerable interval, another observation was recently executed on February 21, 2024. In total, there have been 14 observations, with 10 of them being long-duration observations ($>$ 30\ ks). In this paper, the observation data we used are from January 17, 2004, with an exposure time of 59.7\ ks, and the observation number is ObsID 0205590301. We preprocess the observation data using the Scientific Analysis Software (SAS, version 21.0.0) provided by the \emph{XMM-Newton} Science Operations Center, and extract the required scientific data using the \emph{evselect} tool according to the standard procedure.

During the data analysis, we selected a circular region of interest (ROI) centered at Right Ascension (RA)~$=$3$^h$33$^m$36$^s$.4, Declination (Dec.)~$=$ $-$36$^{\circ}$08$^{\prime}$${26}\farcs{37}$ with a radius of 40 arcseconds, and limited the energy range to events within 0.2 -- 10.0\ keV. We then used the \emph{tabgtigen} tool to filter in time or count rate, generating a good time interval (GTI) file to exclude times when the background flicker rate was high. Next, we used the \emph{evselect} tool to limit the PATTERN of the PN detector to $\leqslant12$ and the PATTERN of the two MOS detectors to $\leqslant4$, set the time bin to 200\ s, and generated a light curve using high-quality scientific data. To eliminate errors caused by background photons, we selected events in a source-free circular ROI with the same radius and without any X-ray sources in the same chips as the source area in order to serve as the background light curve \citep{Zhang2017}. We used \emph{epiclccorr} in SAS to perform background subtraction and correction of detector efficiency. Finally, we combined the light curves from the three detectors (PN + MOS1 + MOS2) and also obtained a combined light curve from the two RGS detectors. The subsequent time series analysis will be based on the combined light curve obtained from the above process. During the spectral analysis, we used the \emph{evselect} tool in the EPIC camera, set the parameter \emph{spectralibinsize} to 200, extracted the spectra from NGC 1365 and the background, and obtained the corresponding response matrix.

Traditional tools for studying time variability include power spectral density (PSD) analysis, LSP, and short-time Fourier transform (STFT) of light curves. PSD and LSP are best suited to searching for strictly periodic signals, and are not designed for analyzing signals with time-varying frequencies. STFT can handle variable frequencies, but it largely depends on the choice of window function, and if chosen improperly, it usually produces significant side effects \citep{Zhao2020}. WWZ has a robust mapping in the frequency-time domain, especially when detecting the time evolution of parameters (such as period, amplitude, and phase) describing periodic and quasi-periodic signals, and shows great advantages as a periodic analysis method \citep{Zhang2020}. Examples of WWZ analysis of the evolution characteristics of QPOs in various celestial systems include \cite{Bravo2014}, \cite{Benkhali2020}, and \cite{Urquhart2022}.

In the present study, we used a version of the WWZ modified with the Morlet mother function \citep{Foster1996}. We conducted a WWZ analysis on the obtained light curve, obtaining three sets of two-dimensional color maps that depict the power spectrum over time. By observing these two-dimensional maps, we can determine whether or not the light curve has a period, and the location of the period. At the same time,  we additionally employed the LSP method for further analysis and to perform comparative checks. The LSP method differs from the principle method used by WWZ. Its advantage is that it can be used to analyze the periodicity in irregular time series by fitting the sine waves in the entire data series using the $\chi^{2}$ statistic. This method reduces the impact of irregular sampling and identifies any periods or quasi-periods that may exist in the data, and it can also calculate their significance.

As shown in Figure \ref{one}, a very strong peak appears at (2.19 $\pm$ 0.08)$\ \times\ $10$^{-4}$\ Hz (4566\ s, error represents full width at half maxima). We further calculated the confidence level that this signal is generated by random noise. First, $10^6$ artificial light curves were generated based on the power spectral density (PSD) and the probability density function of the variations observed in the EPIC light curve. To determine the best-fitting PSD, we used a bending power law plus a constant function to model the PSD of the light curve, and then used the method provided by \citet{Emmanoulopoulos2013} to obtain artificial light curves and evaluate the confidence curves displayed in the lower right panel of Figure \ref{one}. The black solid line represents the result of the WWZ power spectrum averaged over time, and the red solid line represents the result calculated using the LSP method. The figure includes the 3.0, 3.5, and 4.0\ $\sigma$ confidence curves, which were calculated using the method by \citet{Emmanoulopoulos2013}. After accurately calculating the number of tests, we obtained the final confidence level of the QPO as 3.6\ $\sigma$.

In addition, we also used the method proposed by \cite{baluev2008} to calculate the $4.0\ \sigma$ confidence level. Represented by the orange dotted line in Figure \ref {one}, you can see that the result is significantly more than 4.0\ $\sigma$. However, this method does not take into account the influence of red noise, and the results calculated using this method are somewhat controversial for observations that are greatly affected by red noise. Therefore, we can conclude that the QPO discovered in this study has a high level of confidence. In addition to a very obvious peak at a frequency of $2.19 \times 10^{-4}\ {\rm Hz}$ in Figure \ref{one}, there seems to be signals at frequencies of $2.66 \times 10^{-4}\ {\rm Hz}$ and $5.46 \times 10^{-4}\ {\rm Hz}$ in the first $\sim$20 ksec. We pointed out in \cite{Zhang2023} that QPO signals do not persist indefinitely, but rather have a lifespan. This signal may occur at the end of a QPO signal before observation, or it may be a false signal that we cannot make judgments about based on existing information. We can therefore only calculate the confidence level of its signal based on limited information. In this case, we rule out the possibility that these are real signals.

The entire time span of the EPIC light curve was folded with a period of 4566\ s, and the result is shown in Figure \ref{two}. The error of each point is calculated based on the standard deviation ($68.3\ \%$) of the mean values of each phase interval. For clarity, we plotted two periods. The average count rate is $\sim$ 3.07 counts/s. From the figure, it can be seen that the amplitude of the X-ray flux changes significantly with the phase.

In order to more accurately distinguish the energy band from which the QPO originates, we divided 0.2 --\ 10 keV into four parts: 0.2 -- 1.0\ keV, 1.0 -- 2.0\ keV, 2.0 -- 6.0\ keV, and 6.0 -- 10.0\ keV. For different energy bands, we obtained the light curves using the same parameters and data processing methods. Their WWZ results (Figure \ref {three}) show that there was no QPO signal detected in the 0.2 -- 1.0\ keV energy band, and only the hint of a signal at 1.0 -- 2.0\ keV, and the main contribution of the QPO signal comes from the 2.0 -- 10.0\ keV range. Similarly, to more clearly describe this difference, we calculated the confidence level for the 2.0 -- 10.0\ keV energy band. Through $10^6$ artificial light curves, we obtained a confidence level of 3.9\ $\sigma$.

\subsection{Time-averaged spectral analysis}

To study the spectral properties of NGC 1365 more accurately, XSPEC (Version 12.13.1n, \citealt{Arnaud1996}) was employed for the spectral analysis. The energy range we used is from 1.0 to 10.0\ keV. As a starting point, we used a 
simple  initial model consisting of three distinct components related to AGNs: $TBabs \times (powerlaw+pexrav+apec)$ \citep{Nardini2015}, to simultaneously fit the spectra from three EPIC cameras. In this model, TBabs is the Tuebingen-Boulder interstellar medium absorption model, representing the galactic absorption of NGC 1365, fixed at $N_{\rm H}=1.34 \times 10^{20}$ cm$^{-2}$ \citep{Kalberla2005}.  We used a pexrav model \citep{Magdziarz1995} to describe the
part reflected by the irradiated material; this model has solar abundance, an inclination angle of 45$^{\circ}$, and variable intensity. It also includes a soft X-ray power law, interpreted as a dim, scattered AGN continuum. At the same time, to check whether or not a part of the emission line can be attributed to optically thin and collisionally ionized plasma, which is an important component of the \emph{XMM-Newton/RGS} spectrum \citep{Guainazzi2009}, we added an \emph{apec} thermal component with solar abundance \citep{Smith2001}. We used the above model as the initial model for fitting.

The fitting result of this model shows a significant negative deviation at 6.7 -- 8.3\ keV, a phenomenon that has been discussed in several previous papers \citep{Risaliti2005, Risaliti2009}. This is because there are four absorption lines in the X-ray spectrum of NGC 1365 in this range. Through spectral analysis, we identified these features as the $K\alpha$ and $K\beta$ lines of Fe \romannumeral 25 \ and Fe \romannumeral 26. Specifically, the corresponding energies of Fe \romannumeral 25\ $K\alpha$ and Fe \romannumeral 26 \ $K\beta$ are 6.697\ keV and 6.966\ keV, respectively, while those of Fe \romannumeral 25\ $K\alpha$ and Fe \romannumeral 26 \ $K\beta$ are 7.880\ keV and 8.268\ keV \citep{Risaliti2005}. Therefore, we multiplied two \emph{gabs} models \citep{Maitra2018, Klochkov2011} to describe these absorption features.

First, we fixed the \emph{LineE} of the two \emph{gabs} models at 6.832\ keV and 8.074\ keV, respectively, and released them after the other parameters were well fitted. In the end, the \emph{LineE} of the two gabs models were fitted at 6.837\ keV and 7.988\ keV, respectively. The new fit (Model B) brought significant improvement, with a goodness of fit of $\chi^{2}/\nu =1.27$. However, when we checked the fitting data, we found that the fitting result of the \emph{pexrav} model had a large deviation, with a negative photon index power and a reflection scaling parameter very close to zero. After discussion, it was determined that the reflection model was not suitable and was therefore removed. NGC 1365 is considered to show partially obscured X-ray absorption \citep{Risaliti2009, Maiolino2010}, and so we introduced the partially obscured model \emph{TBpcf} \citep{Swain2023} to describe the absorption and multiplied it with the power-law component.

The model is referred to as Model C: $Tbabs \times gabs \times gabs \times (apec+tbpcf \times powerlaw)$. In this model, the parameters of the gabs model first remain the same as the original ones. For the parameters of the \emph{TBpcf} model, we refer to the findings of \citet{Swain2023}, and fix the partial shielding fraction at 0.8. After obtaining the preliminary fitting results, we release the above parameters for further fitting, and obtain a better fit. The goodness of fit of the overall statistical data is $\chi^{2}/\nu=1.26$. After carefully observing the residual plot and logarithmic data plot, we found that there is an emission line near 6.4\ keV, and the distribution of residuals near 8.0\ keV is uneven. Therefore, we introduced two more gabs models and a {Gaussian} emission line model \citep{Brenneman2013}.

\begin{figure}[!h]
\centering
\includegraphics[scale=0.53]{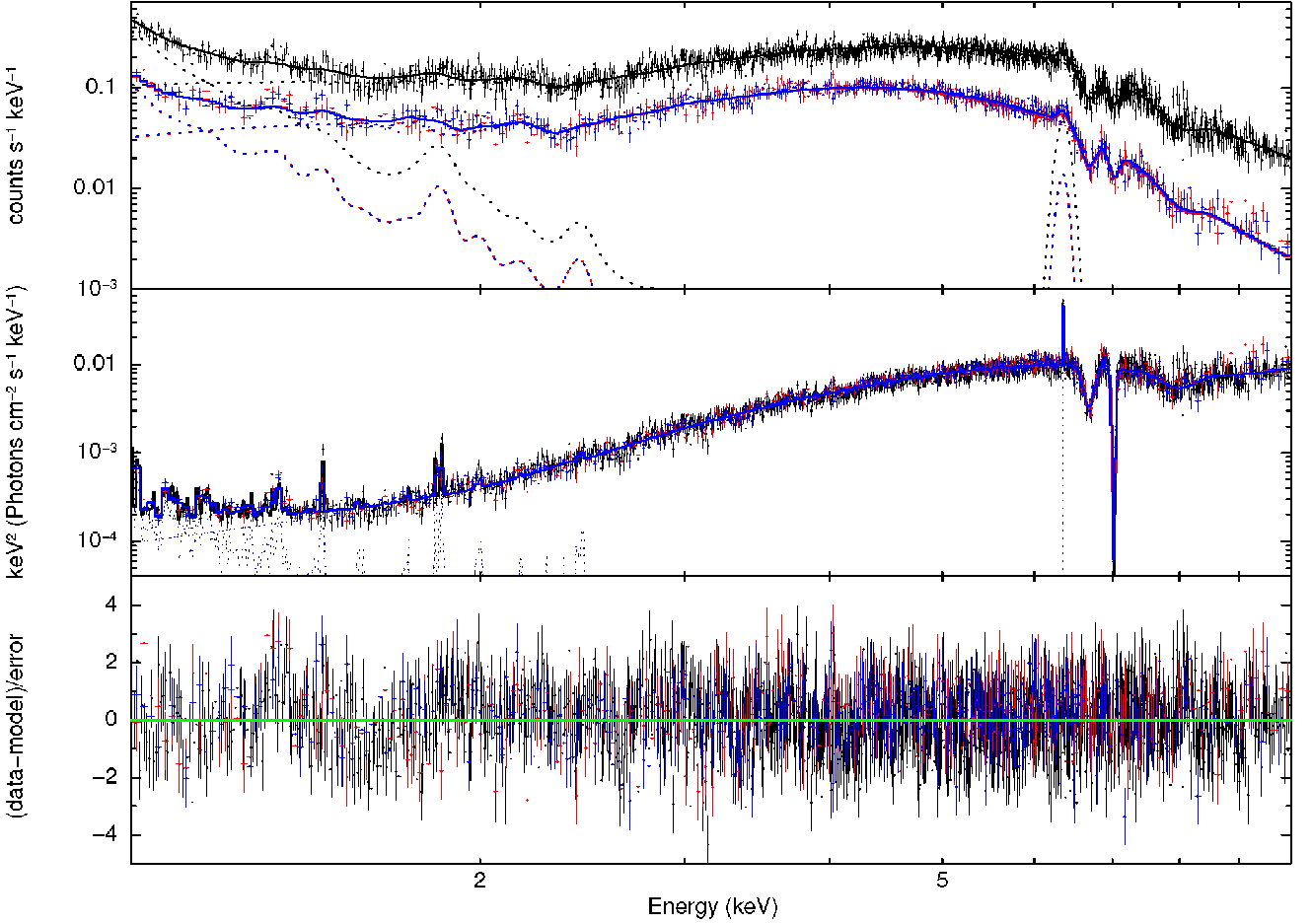}
\caption{
\emph{XMM-Newton} spectrum of NGC 1365 (ObsID 0205590301). Top and middle panel are the logarithmic data and eeuf diagram. Black, red, and blue represent the data of PN, MOS1, and MOS2, respectively. The best-fitting model of Model D and the residuals ([data–model]/error) are shown in the lower panel. 
}
\label{four}
\end{figure}

Therefore, we obtained the final fitting model (Model D): $Tbabs \times gabs \times gabs \times gabs \times gabs \times (gaussian+apec+tbpcf \times powerlaw)$. Based on the original fitting model C, we first fixed the LineE of the Gaussian at 6.40\ keV for fitting, then released it, and finally fitted at 6.35\ keV. This model provides a good fit, with an overall statistic of $\chi^{2}/\nu =1.10$. We compared the results we obtained with the fitting data from the studies of \citet{Risaliti2005}, \citet{Nardini2015}, and \citet{Swain2023}, and the results are somewhat consistent. The best-fitting parameters and related errors are shown in Table \ref{tab1}. The spectrum of the best-fitting model is displayed in the middle panel of Figure \ref{four}; the top panel shows the logarithmic data  plot and the bottom panel shows the residuals.

\begin{table}[!ht]
    \centering
    \caption{Best-fitting spectral parameters derived from Model D.}
    \begin{tabular}{lll}
    \hline
        \textbf{Component} & \textbf{Parameter} & \textbf{Data} \\ 
    \hline
         & LineE & $6.682_{-0.005}^{+0.010}$ \\[1ex]
        Gabs1~ & Sigma & $0.073_{-0.016}^{+0.018}$  \\[1ex]
        ~ & Strength & $0.204_{-0.018}^{+0.019}$ \\ [1ex]
         & LineE & $7.024_{-0.012}^{+0.010}$ \\[1ex]
        ~Gabs2 & Sigma & $0.022_{-0.003}^{+0.003}$  \\[1ex]
        ~ & Strength & $0.321_{-0.099}^{+0.166}$ \\ [1ex]
         & LineE & $7.990_{-0.053}^{+0.051}$ \\[1ex]
        Gabs3~ & Sigma & $0.262_{-0.067}^{+0.074}$  \\[1ex]
        ~ & Strength & $0.260_{-0.072}^{+0.087}$ \\ [1ex]
         & LineE & $9.549_{-0.289}^{+0.433}$ \\[1ex]
        Gabs4~ & Sigma & $2.092_{-0.279}^{+0.381}$  \\[1ex]
        ~ & Strength & $6.984_{-1.652}^{+2.721}$  \\  [1ex]
         & LineE & $6.350_{-0.011}^{+0.022}$ \\[1ex]
        Gaussian & Sigma ($ \times 10^{-4}$) & $7.866_{-1.255}^{+3.224}$ \\[1ex]
        ~ & Norm ($ \times 10^{-5}$) & $2.007_{-0.431}^{+0.520}$ \\ [1ex]
        Apec & kT & $0.791_{-0.050}^{+0.044}$ \\[1ex]
        ~ & Norm ($ \times 10^{-4}$) & $1.470_{-0.181}^{+0.234}$ \\ [1ex]
        TBpcf & $N_H$ ($ \times 10^{22}$) & $7.719_{-0.395}^{+0.394}$ \\[1ex]
        ~ & pcf & $0.863_{-0.029}^{+0.023}$ \\ [1ex]
        Powerlaw & $\Gamma$ & $0.293_{-0.220}^{+0.191}$ \\[1ex]
        ~ & Norm ($ \times 10^{-4}$) & $7.231_{-1.719}^{+2.109}$ \\ [1ex]
        \hline
        Reduced & $\chi^{2}/\nu$ & 1.10 \\
        \hline
    \end{tabular}
        \caption*{Notes. Errors are 90\% confidence limits for one parameter. The units of the parameters in the table are the default model parameters used by XSPEC.}
    \label{tab1}
\end{table}

\begin{figure}[!h]
        \centering
        \includegraphics[scale=0.48]{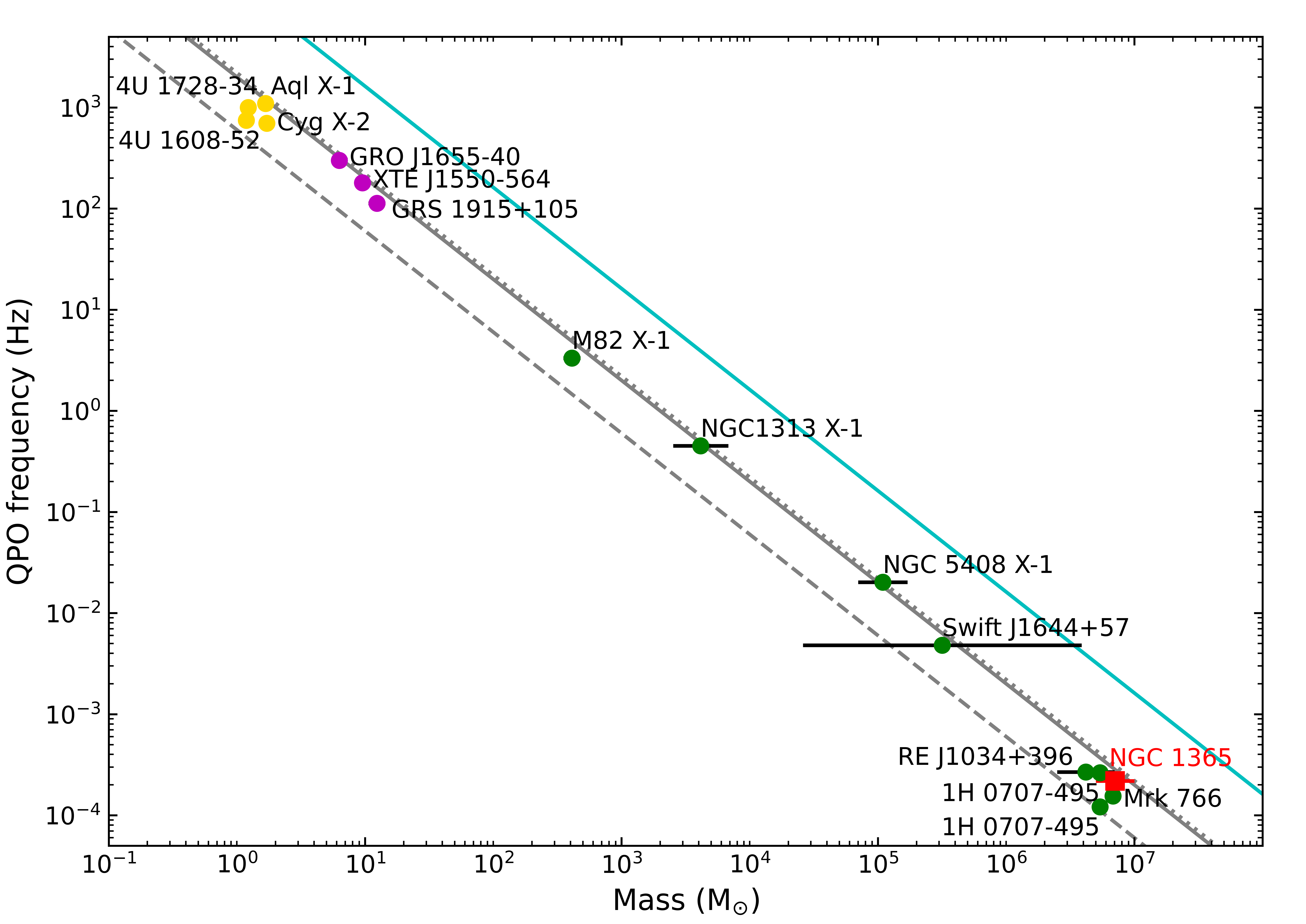}
        \caption{Correlation between source mass and QPO frequency. The yellow points in the upper left corner represent HF QPO events observed in neutron stars, while the purple points correspond to HF QPO events in Galactic microquasar BHs. The remaining points denote HF QPO events reported in supermassive BHs. The newly detected QPO signal in NGC 1365 is marked with a red square. The three gray lines represent the relationships suggested by \citet{Remillard2006} and \citet{Kluzniak2002}. The blue line is based on the maximum allowed orbital frequency proposed by \citet{Goluchova2019}, indicating that QPO signals with frequencies above the innermost stable circular orbit would not exist in the region above and to the right of this line. For further information, please refer to the studies by \citet{Zhou2010, Zhou2015} and \citet{Zhang2017, Zhang2018}.
        }
        \label{five}
\end{figure}

\section{Summary and discussion}
\label{Results}
In this study, we analyzed observational data from the \emph{XMM-Newton} X-ray telescope for the Seyfert galaxy NGC 1365 obtained in January 2004. By employing the WWZ method to construct the time-averaged power spectrum and analyzing it with the LSP method, we identified a QPO signal with a period of 4566 s and a confidence level of 3.6\ $\sigma$ throughout the observation period. However, we did not detect this signal in other \emph{XMM-Newton} observational data. Therefore, we believe that this signal, like the QPO signals found in Mrk 766 \citep{Zhang2017}, 1H 0707$-$495 \citep{Pan2016, Zhang2018}, and Swift J1644$+$57 \citep{Zhou2015}, is a transient periodic signal. 

At present, the origin of QPOs remains a topic of debate. However, existing research generally suggests that QPO signals may be generated by several mechanisms, such as pulsating accretion near the Eddington limit, instabilities of the inner accretion disk, X-ray hotspots orbiting the BH, and disk oscillations and precession (see \citealt{Syunyaev1973, Bardeen1975, Guilbertet1983, Mukhopadhyay2003, Li2003, Remillard2006, Gangopadhyay2012}). The detailed explanatory models include the resonance model \citep{Abramowicz2001}, the relativistic precession model \citep{Stella1999}, the acoustic oscillation model \citep{Rezzolla2003}, the accretion--ejection instability model \citep{Tagger1999}, and disco-seismic models assuming thin disk oscillations \citep{Wagoner1999}.

At the same time, these models also predict the relationship between $f_{\rm QPO}$ and $M_{\rm BH}$. The original diagram and associated lines were first proposed, drawn, and discussed in the works of \citet{Abramowicz2004} and \citet{Torok2005}. In the $f_{\rm QPO}-M_{\rm BH}$ correlation diagram, we first display some of the previous research \citep{ Kluzniak2002, Remillard2006, Zhou2010, Zhou2015, Pan2016}, as shown in Figure \ref{five}. The QPO frequencies found in RE J1034+396 \citep{Gierlinski2008} and 1H 0707$-$495 \citep{Pan2016, Zhang2018} have been considered to be HF QPOs. The QPO frequency of NGC 1365 and the mass of the central BH are plotted on the graph, demonstrating a high level of consistency. This suggests that the discovered QPO may also be classified as HF QPO. The mass data come from the calculation obtained by \citet{Fazeli2019}: the mass of the central BH is in the range of (5 -- 10) $ \times 10^{6}\ M_{\odot}$. The origin of QPO in BH X-ray binaries and Seyfert galaxies is still unclear, and our results can provide more information for this research.

We attempted to fit the energy spectrum ranging from 0.2 to 10.0\ keV, but regardless of the model employed, we were unable to achieve a satisfactory fit. Consequently, we calculated the energy spectra for all observations of NGC 1365, and the results indicate that they exhibit minimal variance in the 0.2 -- 1.0\ keV range. None of them exhibit QPO signals, which leads us to believe that the processes generating QPO signals predominantly emit hard X-ray photons with energies exceeding 1.0\ keV. After excluding photons without QPO signals, the signal's confidence level was enhanced, reaching 3.9\ $\sigma$. We did not observe a situation where QPOs only appeared in the high-energy band in Mrk 766 \citep{Zhang2017} and 1H 0707-495 \citep{Zhang2018}. This scenario likely indicates that the region where QPOs occur primarily emits high-energy X-ray radiation.

The X-ray data in the 1.0 -- 10.0\ keV band were fitted with the \emph{apec} model primarily for the soft band, and the \emph{TBpcf} model  --- which accounts for partial obscuration --- was used in conjunction with a power law to describe the hard band. The fitting results of the spectrum show that the plasma temperature reaches 0.791\ keV, which is close to the results obtained in \citet{Swain2023}. Also, NGC 1365 has multiple absorption lines and emission lines in the high-energy band (> 6.0\ keV), which has been studied many times before \citep{Risaliti2005, Risaliti2009, Nardini2015, Guainazzi2009}, and these emission lines also show puzzling asymmetry \citep{Lena2016}. At the same time, we applied the same model to perform spectral fitting again for the peak and valley regions in the light curve, and compared the model fitting results of the peaks and valleys. The analysis shows that there is no significant difference between the best-fitting data of the peaks and valleys, indicating that the QPO phenomenon is not caused by the spectrum. Considering the causes of other HF QPOs \citep{Abramowicz2001}, we believe that the QPO discovered in this study may also be caused by the resonance mechanism.

At the same time, the gas around the BH causes the absorption or reflection body to show strong heterogeneity, and so the reflection intensity of the gas with the same column density measured by absorption is systematically stronger than expected, and there is partial obscuration \citep{Risaliti2009, Maiolino2010}; therefore, the $N_{\rm H}$ varies greatly under different observation conditions. Unlike standard Seyfert galaxies, such as NGC 2617, the unstable behavior of NGC 1365 below 10.0\ keV makes it difficult to obtain strictly consistent information from different observation times, which also means we cannot come to a unified conclusion on the classification of NGC 1365. The above complex properties mean that the spectrum of NGC 1365 cannot be summed up in a simple description. Therefore, more observations  of NGC 1365 with high spatial resolution and high spectral resolution would help to study the subplasma-scale environmental properties of active galaxies in depth.

\begin{acknowledgements}
This work is supported by the National Natural Science Foundation of China under grants 12403052, 12203029, 12373030, 12233002, U2031205, 2021YFA0718500 and 2023AFB577. The data used in this paper were obtained from the \emph{XMM-Newton}, and we express our gratitude.
\end{acknowledgements}

\bibliographystyle{aa}
\bibliography{myBiblio.bib}

\label{lastpage}
\end{document}